\documentclass[10pt]{article}
\usepackage{amssymb}
\usepackage{amsmath}
\usepackage{graphicx}
\usepackage{fancyhdr}
\usepackage{cite}

\begin{document}

\newcommand{\bin}[2]{\left(\begin{array}{c}\!#1\!\\\!#2\!\end{array}\right)}

\huge

\begin{center}
Modeling penetrating collisions in the standard line broadening impact theory for hydrogen
\end{center}

\vspace{0.5cm}

\large

\begin{center}
Jean-Christophe Pain\footnote{jean-christophe.pain@cea.fr} and Franck Gilleron
\end{center}

\normalsize

\begin{center}
CEA, DAM, DIF, F-91297 Arpajon, France
\end{center}

\vspace{0.5cm}


\begin{abstract}
Spectral lines emitted by plasmas provide information about the thermodynamic conditions, the degree of randomness or the interactions prevailing in the medium. Collisions by plasma electrons penetrating the extent of bound-electron wavefunctions is important at high density, where short-range interactions become dominant. Such collisions are usually not taken into account properly in the standard lineshape theory, assuming long-range dipole approximation. The formalism of penetrating collisions for hydrogen relies on the introduction of a family of integrals calculated using a recursion relation. In this work, we show that such integrals can be expressed analytically, as a finite sum involving binomial coefficients and modified Bessel functions of the third kind. The explicit expression enabled us to obtain a simple approximate analytical form for the collision operator, making numerical implementation and physical interpretation easier. We also propose simple analytical forms of coefficients and integrals important for the modeling of penetrating collisions.
\end{abstract}

\section{Introduction}

Emission spectroscopy is a powerful tool to infer plasma conditions, for instance to perform electron temperature and density diagnostics. In the so-called ``standard line shape theory'', electrons are modeled in the impact theory, which relies on a classical path trajectory and often resorts to a second-order perturbative treatment of the self-energy, and the ions are described in the quasi-static approximation. For both electrons and ions, the emitter-perturber interaction is often assumed to be dipolar only. One of the main weaknesses of the standard electron treatment concerns the so-called strong collisions, \emph{i.e.} collisions associated to small impact parameters (and small electron velocities), for which the dipole/quadrupole interaction used for the interaction is questionable due to penetration by the perturbing electrons into the atomic (bound-state) wavefunction extent. A few years ago, it was suggested \cite{GRIEM97,ALEXIOU06} that penetration was likely to be more important than thought because the standard cutoff $n^2a_0/Z$ ($n$ is the principal quantum number $a_0$ the Bohr radius and $Z$ the atomic number) representing the wavefunction extent was too optimistic. In addition, penetration softens the interaction so that perturbation theory remains valid, even for collisions previously considered as strong \cite{ALEXIOU05}. The standard theory expression for the self-energy collision operator $\Phi$ is

\begin{equation}\label{colgen}
\langle\alpha\beta|\Phi|\alpha'\beta'\rangle=\sum_{\alpha''}\mathbf{r}_{\alpha\alpha''}.\mathbf{r}_{\alpha''\alpha'}\phi_{\alpha\alpha'',\alpha''\alpha'}+\sum_{\beta''}\mathbf{r}_{\beta'\beta''}.\mathbf{r}_{\beta''\beta}\phi_{\beta'\beta'',\beta''\beta}-\mathbf{r}_{\alpha\alpha'}.\mathbf{r}_{\beta'\beta}\phi^{\mathrm{int}}_{\alpha\alpha',\beta'\beta},
\end{equation}

\noindent where $\alpha$ and $\alpha'$ are upper level states, $\alpha''$ is a state perturbing the upper level states and $\mathbf{r}_{ij}$ are matrix elements of the position operator. By this, it is meant that a collision with a plasma electron has a nonnegligible probability amplitude to cause a transition $\alpha\rightarrow\alpha''$. Similarly, $\beta$ and $\beta'$ are lower level states and $\beta''$ perturbs them. In this work, which will deal with hydrogen only, we employ the no-quenching approximation, \emph {i.e.} $\alpha,\alpha',\alpha''$ have the upper-level principal quantum number ($n_{\alpha}=n_{\alpha'}=n_{\alpha''}$) and $\beta,\beta',\beta''$ have the lower-level principal quantum number ($n_{\beta}=n_{\beta'}=n_{\beta''}$). The quantity $\phi$ of Eq. (\ref{colgen}) is essentially the velocity integrated complex function of standard theory and $\phi^{\mathrm{int}}$ the inteference term. In the penetrating standard theory, it reads, choosing explicitely a straight line trajectory $\mathbf{R}(t)=\mathbf{\rho}+\mathbf{v}t$: 

\begin{equation}\label{coll}
\phi_{\alpha\alpha'',\alpha''\alpha'}=-\frac{\pi n_e}{3}\left(\frac{e^2}{4\pi\epsilon_0\hbar}\right)^2\int vf(v)dv\int\rho I\left(\rho,v;n_{\alpha},\ell_{\alpha},n_{\alpha''},\ell_{\alpha''}\right)\nonumber I\left(\rho,v;n_{\alpha''},\ell_{\alpha''},n_{\alpha'},\ell_{\alpha'}\right)d\rho,
\end{equation}

\noindent where $e$ is the electron charge, $m$ the electron mass, $n_e$ the electron density and $n_i$ and $\ell_i$ are respectively the principal and orbital quantum numbers. The quantity $f(v)$ represents the velocity ($v$) distribution of the perturber, $\rho$ the impact parameter and

\begin{equation}\label{i}
I(\rho,v;n,\ell,n',\ell')=\rho\int_{-\infty}^{\infty}\frac{C_1\left(n,\ell,n',\ell';\sqrt{\rho^2+v^2t^2}\right)}{\left(\rho^2+v^2t^2\right)^{3/2}}dt.
\end{equation}

\noindent The integral $I$ in Eq. (\ref{i}) essentially includes the atomic-collision physics and $C_1$ is a factor accounting exactly for penetration in the dipolar approximation. It is a particular case of $C_{\lambda}$ ($\lambda$ is actually the multipolarity) for which we derived new interesting relations (see Appendix). The standard behavior is recovered if $C_{\lambda}=1$ (no penetration) and in that case $I=2/(\rho v)$. Penetration ``softens'' the interaction in the sense that it tends to reduce the broadening, at least for isolated lines \cite{ALEXIOU01}. However, Alexiou has shown that in some cases, especially for strong coupling conditions, penetrating collisions can enhance the broadening \cite{ALEXIOU17b}, when small impact parameters are involved and when the shielding length becomes of the same order as the wavefunction extent, for instance in the case of line merging \cite{INGLIS39}. The calculation of $I$ is tedious, and Alexiou and Poqu\'erusse have shown that it can be expressed as

\begin{equation}\label{col1}
I=\frac{2}{\rho v}\left[1-\Delta(b)\right]
\end{equation}

\noindent with $b=a\rho$ and 

\begin{equation}\label{del}
\Delta(b)=\sum_{i=0}^{2n+\lambda}s_ib^iF_{i-2}(b).
\end{equation}

The coefficients $s_i$, which are rapidly decreasing functions of $i$, can be computed exactly and are provided in the Appendix of Ref. \cite{ALEXIOU05}. They read

\begin{equation}
s_i=D^{-1}\left[\sum_{j=\max(i,\lambda+\ell+\ell'+2)}^{2n+\lambda}\frac{c_jj!}{i!}-\theta(i-2\lambda-1)c_j\frac{(j-2\lambda-1)!}{(i-2\lambda-1)!}\right],
\end{equation}

\noindent where the coefficients  $D$ and $c_j$ are given in the Appendix and $\theta(j)=1$ if $j\ge 0$ and 0 otherwise. The final collision operator (see Eq. (\ref{coll})) is then reduced to one-dimensional quadratures:

\begin{equation}\label{col2}
\phi=-\frac{4\pi e^4n_e}{3\left(4\pi\epsilon_0\hbar\right)^2}\sqrt{\frac{2m}{\pi k_BT}}\int_0^{b_{\mathrm{max}}}\frac{db}{b}\left[1-\Delta\left(b;n_{\alpha},\ell_{\alpha},n_{\alpha},\ell_{\alpha''}\right)\right]\left[1-\Delta\left(b;n_{\alpha},\ell_{\alpha''},n_{\alpha},\ell_{\alpha'}\right)\right]\nonumber\\
\end{equation}

\noindent for the direct term associated to the upper states, where $\alpha$ refers to states of the upper levels and $b_{\mathrm{max}}=a\rho_{\mathrm{max}}$, with $a=2/\left(n_{\alpha}a_0\right)$. The direct term associated to the lower states is obtained by replacing the indices $\alpha$ by $\beta$ in Eq. (\ref{col2}). For the interference term, we have

\begin{equation}\label{col3}
\phi^{\mathrm{int}}=-\frac{8\pi e^4n_e}{3\left(4\pi\epsilon_0\hbar\right)^2}\sqrt{\frac{2m}{\pi k_BT}}\int_0^{b_{\mathrm{max}}}\frac{db}{b}\left[1-\Delta\left(b;n_{\alpha},\ell_{\alpha},n_{\alpha},\ell_{\alpha'}\right)\right]\left[1-\Delta\left(\frac{n_{\alpha}b}{n_{\beta}};n_{\beta},\ell_{\beta'},n_{\beta},\ell_{\beta}\right)\right],\nonumber\\
\end{equation}

\noindent where $\alpha$ refers to states of the upper and $\beta$ to the states of the lower levels. $b_{\mathrm{max}}$ is the maximum value of $b$ for the upper level.

It turns out that expression of $\Delta(b)$ in Eq. (\ref{del}) involves integrals of the kind

\begin{equation}\label{fq}
F_q(b)=\int_0^{\infty}e^{-b\cosh(u)}\cosh^q(u)du,
\end{equation}

\noindent for $q=-2, \cdots, 2n+\lambda-2$ and Alexiou and Poqu\'erusse suggested to compute $F_q$ using the four-term recursion relation:

\begin{equation}\label{rec}
F_{q+2}(b)=\frac{(q+1)}{b}F_{q+1}(b)+F_q(b)-\frac{q}{b}F_{q-1}(b),
\end{equation}

\noindent initialized by $F_{-2}(b)=Ki_2(b)$, $F_{-1}(b)=Ki_1(b)$ and $F_0(b)=K_0(b)$, where $Ki_n$ is the Bickley-Naylor function of order $n$ \cite{BICKLEY35} and $K_n$ the modified Bessel function of the third kind of order $n$ \cite{ABRAMOWITZ64}. However, the authors could not find an exact expression solution of Eq. (\ref{rec}) and noticed that $F_q$ can be expressed as $K_0$ and $K_1$ Bessel functions, multiplied by respective coefficients obeying themselves a recursion relation. We found that it is possible to obtain an explicit analytic expression of $F_q$ (see Sec. \ref{sec1}). The asymptotics of the latter quantity is discussed in Sec. \ref{sec2}, and the analytic expression is used to derive an approximate simple and accurate formulation of the collision operator, presented in Sec. \ref{sec3}.

\section{Explicit expression of integral $F_q$}\label{sec1}

Using the relation \cite{GRADSHTEYN80} ($p$ is a positive integer):

\begin{equation}\label{even}
\cosh^{2p}(u)=\frac{1}{2^{2p}}\left\{2\sum_{k=0}^{p-1}\bin{2p}{k}\cosh\left[(2p-2k)u\right]+\bin{2p}{p}\right\}
\end{equation}

\noindent for even powers of $\cosh$ and

\begin{equation}\label{odd}
\cosh^{2p+1}(u)=\frac{1}{2^{2p}}\sum_{k=0}^{p}\bin{2p+1}{k}\cosh\left[(2p-2k+1)u\right]
\end{equation}

\noindent for odd powers of $\cosh$, and the expression of the modified Bessel function of the third kind of order $n$ ($b$ is strictly positive):

\begin{equation}
K_n(b)=\int_0^{\infty}e^{-b\cosh(u)}\cosh(nu)du,
\end{equation}

\noindent we get, inserting expressions (\ref{even}) and (\ref{odd}) into Eq. (\ref{fq}):

\begin{equation}\label{feven}
F_{2p}(b)=\frac{1}{2^{2p}}\left\{2\sum_{k=0}^{p-1}\bin{2p}{k}K_{2p-2k}\left(b\right)+\bin{2p}{p}K_0(b)\right\}
\end{equation}

\noindent for even values and

\begin{equation}\label{fodd}
F_{2p+1}(b)=\frac{1}{2^{2p}}\sum_{k=0}^{p}\bin{2p+1}{k}K_{2p-2k+1}\left(b\right)
\end{equation}

\noindent for odd values. Therefore, $F_q$ can be written as a finite sum of binomial coefficients multiplied by modified Bessel functions of the third kind. $F_q$ contains only even-order Bessel functions if $q$ is even, and only odd-order Bessel functions if $q$ is odd. Equations (\ref{feven}) and (\ref{fodd}) constitute the main results of the present work. We have for instance:

\begin{equation}
F_4(b)=\frac{\bin{4}{2}K_0(b)+2\bin{4}{1}K_2(b)+2\bin{4}{0}K_4(b)}{16}=\frac{3}{8}K_0(b)+\frac{1}{2}K_2(b)+\frac{1}{8}K_4(b)\nonumber\\
\end{equation}

\noindent and

\begin{equation}
F_5(b)=\frac{\bin{5}{2}K_1(b)+\bin{5}{1}K_3(b)+\bin{5}{0}K_5(b)}{16}=\frac{5}{8}K_1(b)+\frac{5}{16}K_3(b)+\frac{1}{16}K_5(b).\nonumber\\
\end{equation}

The same procedure can be applied for quantities 

\begin{eqnarray}
N_k=\int_0^{\infty}e^{\xi\left[1-\epsilon\cosh(u)\right]}\left(\epsilon\cosh(u)-1\right)^kdu
\end{eqnarray}

\noindent used in other publications by the same authors for hydrogenic ions. In that case, which is beyond the scope of the present work (focusing on hydrogen), $\epsilon$ is the eccentricity of the electron trajectory and 

\begin{equation}
\xi=\frac{2Zs}{na_0},
\end{equation}

\noindent where $Z$ is the atomic number, $n$ the principal quantum number of the upper or lower level and 

\begin{equation}
s=\frac{(Z-1)e^2}{4\pi\epsilon_0mv^2}.
\end{equation}

\noindent Indeed, using the binomial expansion

\begin{equation}
\left(\epsilon\cosh(u)-1\right)^k=\sum_{p=0}^k(-1)^ {k-p}\bin{k}{p}\cosh^p(u)\epsilon^{p},
\end{equation}

\noindent we can write

\begin{equation}
N_k=(-1)^ke^{\xi}\sum_{p=0}^k\bin{k}{p}(-\epsilon)^{p}\int_0^{\infty}e^{-\left(\xi\epsilon\right)\cosh(u)}\cosh^p(u)du=(-1)^ke^{\xi}\sum_{p=0}^k(-1)^p\bin{k}{p}\epsilon^{p}F_p\left(\xi\epsilon\right).
\end{equation}

Expressions (\ref{feven}) and (\ref{fodd}) may also be of numerical interest. The clever recursion relation of Alexiou and Poqu\'erusse (Eq. (\ref{rec})) is already efficient numerically, but if the first values are not evaluated with a sufficient accuracy, the error can propagate through the recursion and increase. Moreover, the authors combined the recurrence with asymptotic expressions. Our relations, although they involve special functions, can be calculated accurately, provided that the Bessel functions are computed properly (the literature about the calculation of Bessel function is abundant, see for instance Refs. \cite{TEMME75,GATTO81}).

\section{Asymptotic behaviors}\label{sec2}

The asymptotic forms of Bessel function $K_n$ and $Ki_n$ are respectively \cite{ABRAMOWITZ64}:

\begin{equation}
K_n(z)\approx\sqrt{\frac{\pi}{2z}}e^{-z}\left\{1+\frac{\left(4n^2-1\right)}{8z}+\frac{\left(4n^2-1\right)\left(4n^2-9\right)}{2!(8z)^2}+\frac{\left(4n^2-1\right)\left(4n^2-9\right)\left(4n^2-25\right)}{3!(8z)^3}+\cdots\right\}.
\end{equation}

\noindent and

\begin{equation}
Ki_n(z)\approx\sqrt{\frac{\pi}{2z}}e^{-z}\left[1+\frac{1}{(n-1)!}\sum_{m=1}^{\infty}\frac{(-1)^m}{z^m}\frac{(2m-1)!}{2^{2m-1}(m-1)!}\sum_{k=0}^m\frac{(2k)!(n+m-k-1)!}{8^k(k!)^2(m-k)!}\right].
\end{equation}

Therefore, for large values of $b$, one has, keeping only the first term of the preceeding asymptotic expansions:

\begin{equation}\label{asi}
\Delta(b)\approx\sqrt{\frac{\pi}{2b}}e^{-b}\sum_{i=0}^{n+n'+\lambda}s_ib^i.
\end{equation}

\begin{figure}
\begin{center}
\vspace{1cm}
\includegraphics[width=8cm]{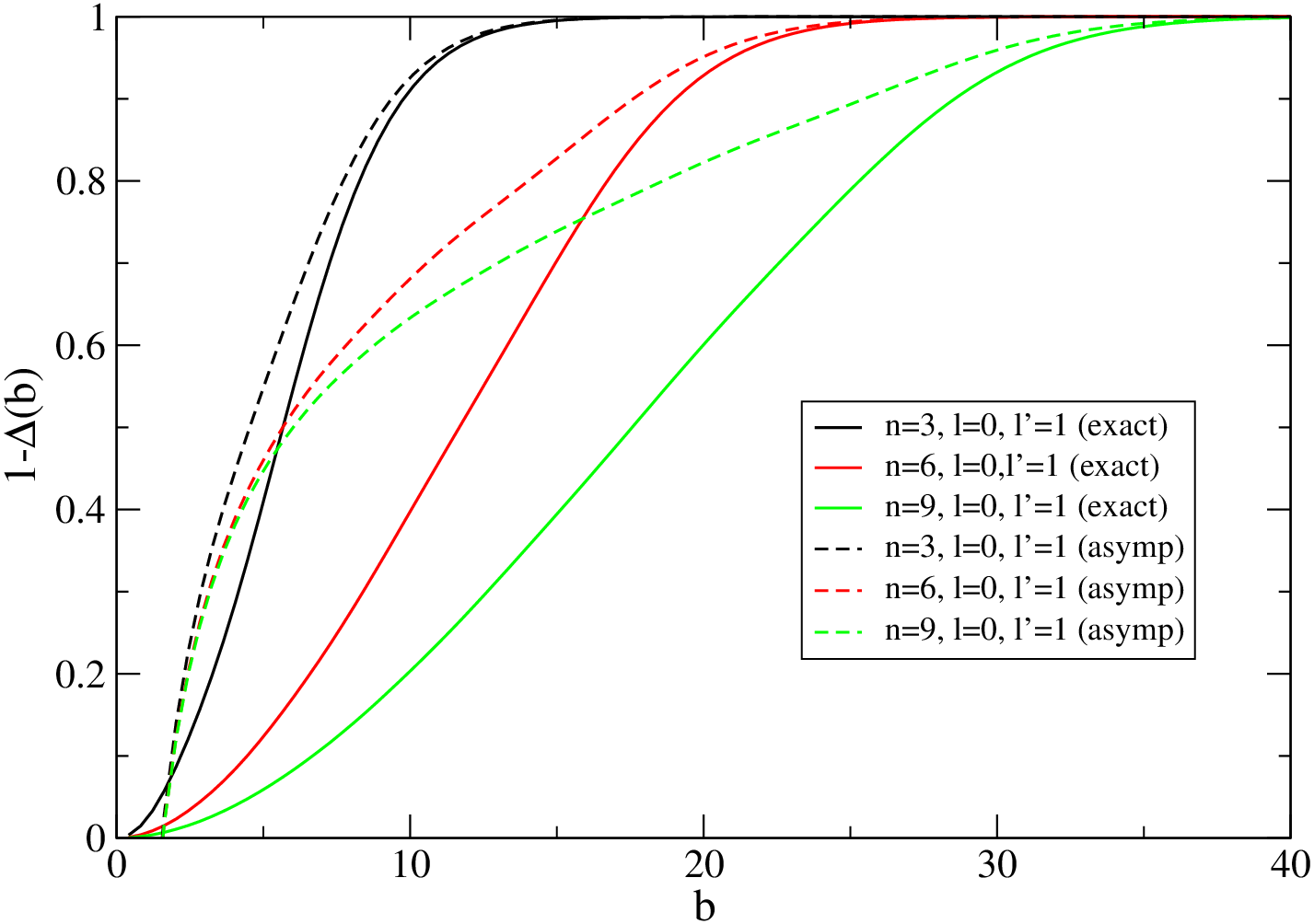}
\end{center}
\caption{Comparison between the exact computation of $\left[1-\Delta(b)\right]$ (Eqs. (\ref{del}), (\ref{feven}) and (\ref{fodd})) and the asymptotic form (\ref{asi}).}\label{fig1}
\end{figure}

Figure \ref{fig1} displays a comparison of the exact computation and the latter asymptotic form. For $n=3$, $\ell=0$ and $\ell'=1$, $\Delta(b=15)$ is very close to the exact value, for $n=6$, $\ell=0$ and $\ell'=1$, it differs from the exact value by about 15 \%, and for $n=9$, $\ell=0$ and $\ell'=1$, it differs from the exact value by 46 \%.

\section{Approximate form of the collision operator}\label{sec3}

We show that our new expression of $F_q(b)$ (Eqs. (\ref{feven}) and (\ref{fodd})) enables one to obtain a simple approximate formula for $\Delta(b)$ which integral gives the collision operator (see Eqs. (\ref{col1}), (\ref{col2}) and (\ref{col3})). Noticing that the quantity $\Delta(b)$ has a half-bell shape (see Fig. \ref{fig1}) with $\Delta(0)=1$ and $\Delta(\infty)=0$, we tried to find an approximation with the function

\begin{equation}\label{app}
\Delta_{\mathrm{app}}\left(b;n,\ell,\ell''\right)=\exp\left[-\frac{b^2}{2\chi_{n,\ell,\ell''}^2}\right],
\end{equation}

\noindent where $\chi_{n,\ell,\ell''}$ can be determined by ensuring the preservation of the zero$^{th}$-order moment (area) of $\Delta(b)$:

\begin{equation}
\chi_{n,\ell,\ell''}=\sqrt{\frac{2}{\pi}}\int_0^{\infty}\Delta(b)db.
\end{equation}

Actually, the quantity $\Delta(b)$ can be expressed without resorting to Bickley-Naylor functions (the first two terms in Eq. (\ref{del}) are respectively $F_{-2}(b)=Ki_2(b)$ and $F_{-1}(b)=Ki_1(b)$ and are rather difficult to handle). Indeed, since

\begin{equation}
Ki_2(b)=b\left[K_1(b)-Ki_1(b)\right],
\end{equation} 

\noindent we get, using $s_0=s_1=1$:

\begin{equation}
\Delta(b)=\sum_{i=2}^{2n+\lambda}s_ib^iF_{i-2}(b)+bK_1(b).
\end{equation}

\noindent Taking into account the fact that

\begin{equation}
\int_0^{\infty}b^iK_n(b)db=2^{i-1}\Gamma\left(\frac{1+i-n}{2}\right)\Gamma\left(\frac{1+i+n}{2}\right),
\end{equation}

\noindent for $i\ge n$, where $\Gamma$ is the usual Gamma function, we can show that, in the dipolar case ($\lambda=1$):

\begin{equation}\label{exactarea}
\int_0^{\infty}\Delta(b)db=\frac{\pi}{4}\sum_{i=0}^{2n+1}s_ii!=\frac{\pi}{8}~\frac{\left[5n^2-\ell_<(\ell_<+2)\right]}{n},
\end{equation}

\noindent where $\ell_<=\min\left(\ell,\ell'\right)$. The final expression on the right-hand side of Eq. (\ref{exactarea}) is exact and has been obtained by a numerical study of the sum over a wide range of parameters $n$, $\ell$ and $\ell'$. As a result, the parameter of the Gaussian approximate function $\Delta_{\mathrm{app}}\left(b\right)$ reads

\begin{equation}
\chi_{n,\ell,\ell'}=\frac{\sqrt{2\pi}}{8}~\frac{\left[5n^2-\ell_<\left(\ell_<+2\right)\right]}{n}.
\end{equation}

The Gaussian function allows one to define roughly the range of impact parameter $\left[0,\rho_c\right]$ for which penetrating collisions are important (\emph{i.e.}, $b$ for which $\Delta(b)>0$). Defining $b_c$ such that $\Delta(b_c)=e^{-4}\approx0.018$, one obtains

\begin{equation}
b_c=\frac{\sqrt{\pi}}{2}\frac{\left[5n^2-\ell_<\left(\ell_<+2\right)\right]}{n},
\end{equation}

\noindent or 

\begin{equation}
\rho_c=a_0\frac{\sqrt{\pi}}{4}\left[5n^2-\ell_<\left(\ell_<+2\right)\right].
\end{equation}

\begin{figure}
\begin{center}
\vspace{1cm}
\includegraphics[width=8cm]{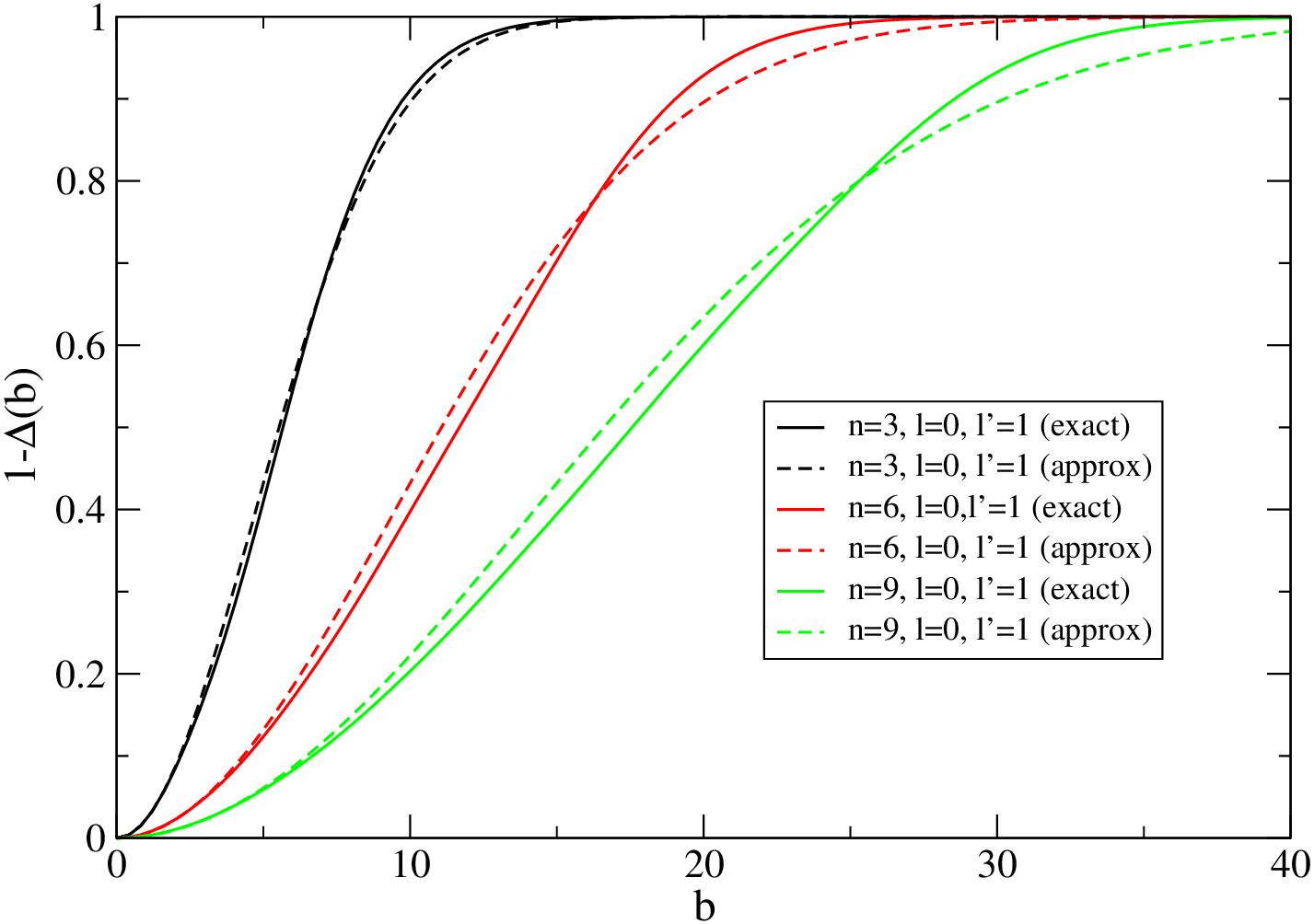}
\end{center}
\caption{Comparison between the exact computation of $\left[1-\Delta(b)\right]$ (Eqs. (\ref{del}), (\ref{feven}) and (\ref{fodd})) and the approximate expression (\ref{app}) for $\ell=0$, $\ell'=1$ and three values of principal quantum number $n$: 3, 6 and 9.}\label{fig2}
\end{figure}

As expected, this corresponds to impact parameters less than roughly the mean radius of shell $n$, \emph{i.e.}, $n^2a_0$. Figure \ref{fig2} displays the comparison between the exact calculation and our approximate expression in three cases $n$=3, 6 and 9 for $\ell=0$ and $\ell'=1$. We can see that the approximate formula is rather accurate, especially for the small values of $n$.

\begin{figure}
\begin{center}
\vspace{1cm}
\includegraphics[width=8cm]{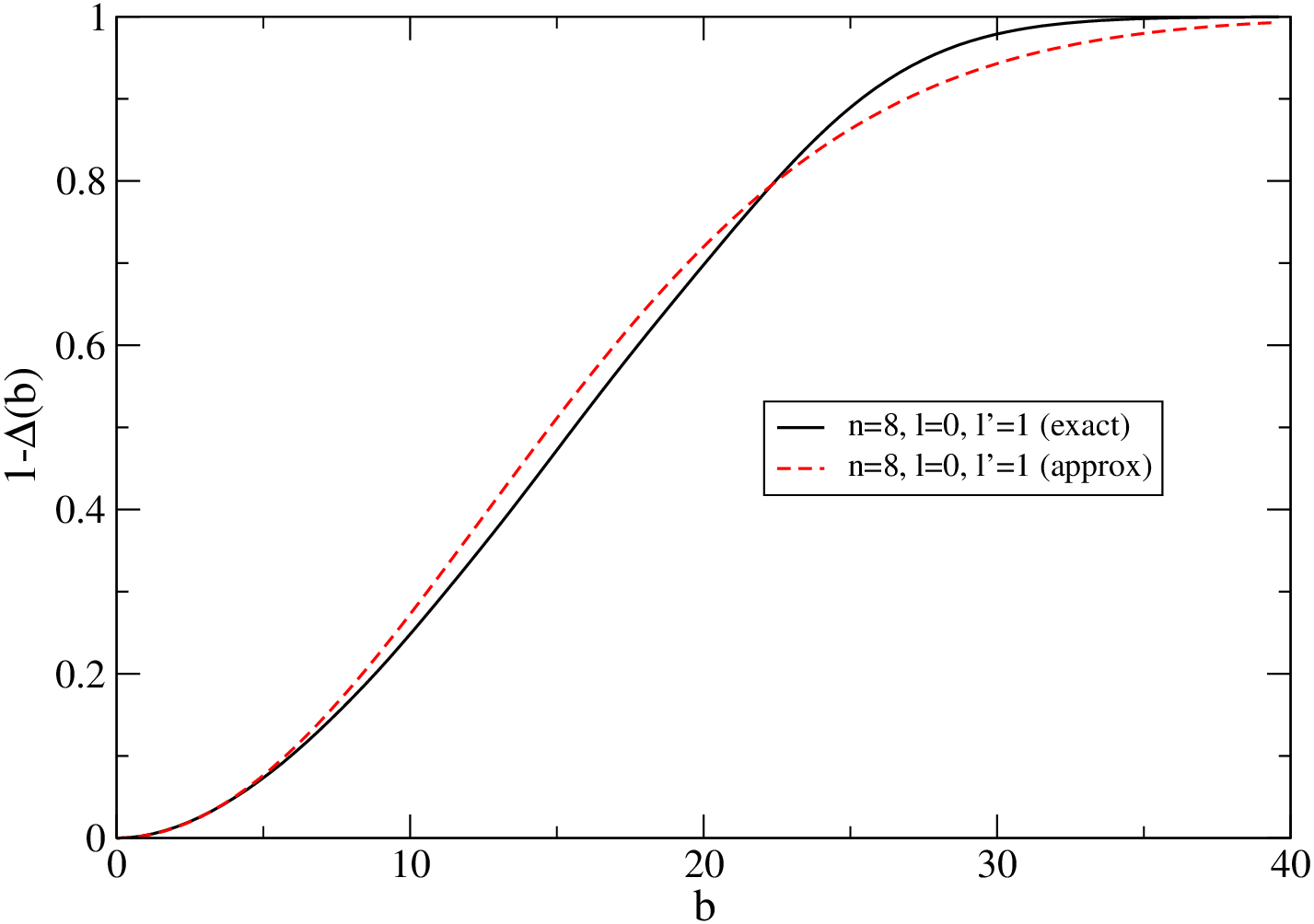}
\end{center}
\caption{Comparison between the exact computation of $\left[1-\Delta(b)\right]$ (see Eqs. (\ref{del}), (\ref{feven}) and (\ref{fodd})) and the approximate expression (\ref{app}) for $n=8$, $\ell=0$ and $\ell'=1$.}\label{fig3}
\end{figure}

\begin{figure}
\begin{center}
\vspace{1cm}
\includegraphics[width=8cm]{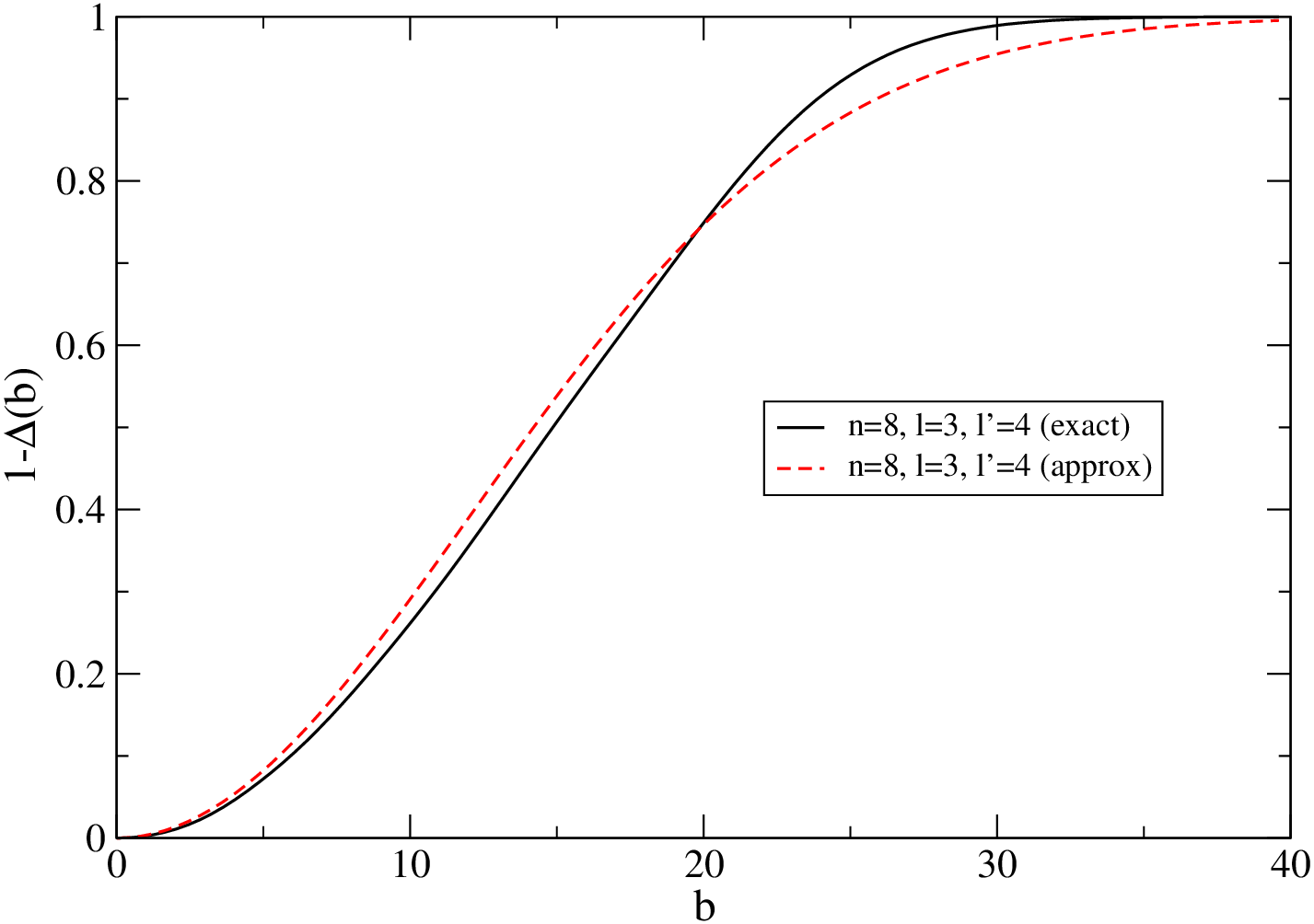}
\end{center}
\caption{Comparison between the exact computation of $\left[1-\Delta(b)\right]$ (see Eqs. (\ref{del}), (\ref{feven}) and (\ref{fodd})) and the approximate expression (\ref{app}) for $n=8$, $\ell=3$ and $\ell'=4$.}\label{fig4}
\end{figure}

\begin{figure}
\begin{center}
\vspace{1cm}
\includegraphics[width=8cm]{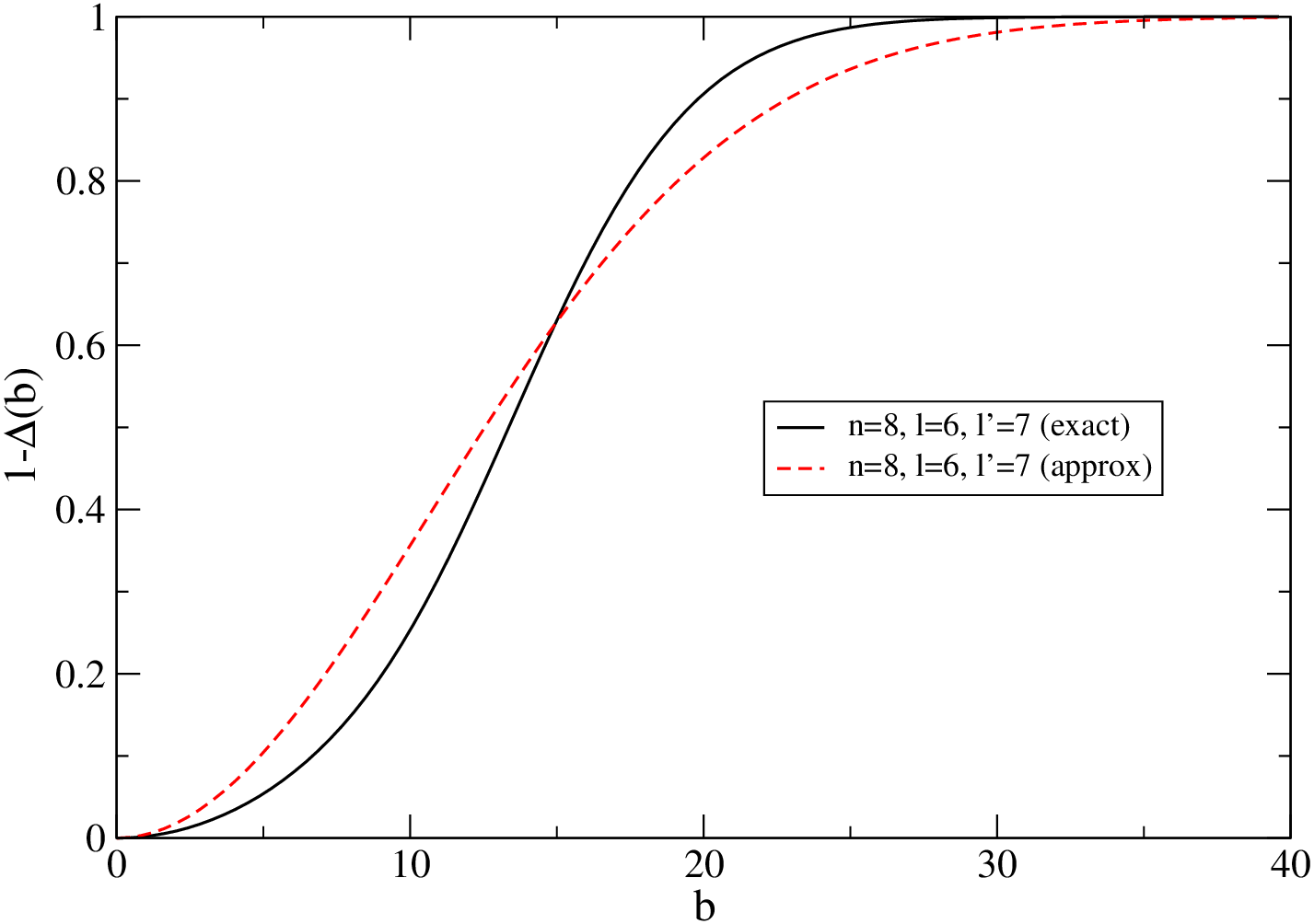}
\end{center}
\caption{Comparison between the exact computation of $\left[1-\Delta(b)\right]$ (see Eqs. (\ref{del}), (\ref{feven}) and (\ref{fodd})) and the approximate expression (\ref{app}) for $n=8$, $\ell=6$ and $\ell'=7$.}\label{fig5}
\end{figure}

It is instructive to fix the principal quantum number $n$ and vary the orbital quantum numbers $\ell$ and $\ell'$. Figures \ref{fig3}, \ref{fig4} and \ref{fig5} show a comparison betwen the exact and approximate forms for $n=8$ in the three cases ($\ell=0$, $\ell'=1$), ($\ell=3$, $\ell'=4$) and ($\ell=6$, $\ell'=7$) respectively. We can see that the accuracy of $\Delta_{\mathrm{app}}(b)$ decreases as $\ell$ increases but its quality remains rather satisfactory.

In fact, a better approximation of $\Delta(b)$ can be obtained using a multi-parameter function (instead of just one for the Gaussian). For instance, one may constrain the approximate function in order to preserve several moments which can be computed exactly through the formula (for $m\geq$ 0):

\begin{equation}
\int_0^{\infty}b^m\Delta(b)db=\frac{\sqrt{\pi}}{2}\frac{\Gamma\left(\frac{m+3}{2}\right)}{\Gamma\left(\frac{m+4}{2}\right)}\sum_{i=0}^{2n+1}s_i(m+i)!.
\end{equation}

Moreover, the formula (\ref{app}) for $\Delta(b)$ enables one to obtain an analytical approximate expression for the collision operator itself (Eqs. (\ref{col2}) and (\ref{col3})). For example, the contribution of the upper-level states $n_{\alpha},\ell_{\alpha},\ell_{\alpha''}$ to the diagonal part of the collision operator reads

\begin{equation}\label{diag}
\phi_{\alpha\alpha'',\alpha''\alpha}=-\frac{4\pi}{3}n_e\left(\frac{e^2}{4\pi\epsilon_0\hbar}\right)^2\sqrt{\frac{2m}{\pi k_BT}}\int_0^{b_{\mathrm{max}}}\frac{\left[1-\Delta\left(b;n_{\alpha},\ell_{\alpha},n_{\alpha},\ell_{\alpha''}\right)\right]^2}{b}db,
\end{equation}

\noindent where $b_{\mathrm{max}}=2\rho_{\mathrm{max}}/\left(n_{\alpha}a_0\right)$ is a cutoff introduced to avoid the logarithmic divergence of the integral at large impact parameters (because $\Delta(b)\rightarrow 0$ when $b\rightarrow\infty$). As for the standard theory, the maximum impact parameter $\rho_{\mathrm{max}}$ is usually chosen to be of the order of the Debye length

\begin{equation}
\lambda_D=\sqrt{\frac{\epsilon_0k_BT}{n_ee^2}}
\end{equation}

\noindent or 1.1$\lambda_D$ (respectively 0.68$\lambda_D$) to account for the single \cite{GRIEM62} (respectively double \cite{CHAPPELL69}) shielded fields in the $S$ matrix. As a result, the Gaussian approximation yields

\begin{equation}\label{diagapp}
\phi_{\alpha\alpha'',\alpha''\alpha}=-\frac{4\pi}{3}n_e\left(\frac{e^2}{4\pi\epsilon_0\hbar}\right)^2\sqrt{\frac{2m}{\pi k_BT}}~\tilde{\phi}_{\alpha\alpha''},
\end{equation}

\noindent with

\begin{equation}
\tilde{\phi}_{\alpha\alpha''}=f\left(\frac{b_{\mathrm{max}}}{\chi_{n_{\alpha},\ell_{\alpha},\ell_{\alpha''}}}\right),
\end{equation}

\noindent the function $f$ being defined as

\begin{equation}\label{f}
f(x)=\frac{\gamma_E}{2}-\frac{1}{2}E_1\left(x^2\right)+E_1\left(\frac{x^2}{2}\right)+\ln\left(\frac{x}{2}\right),
\end{equation}

\noindent where $\gamma_E$ is the Euler-Mascheroni constant \cite{ABRAMOWITZ64} and $E_1$ represents the exponential integral

\begin{equation}
E_1(z)=\int_z^{\infty}\frac{e^{-t}}{t}dt.
\end{equation}

\begin{figure}
\begin{center}
\vspace{1cm}
\includegraphics[width=8cm]{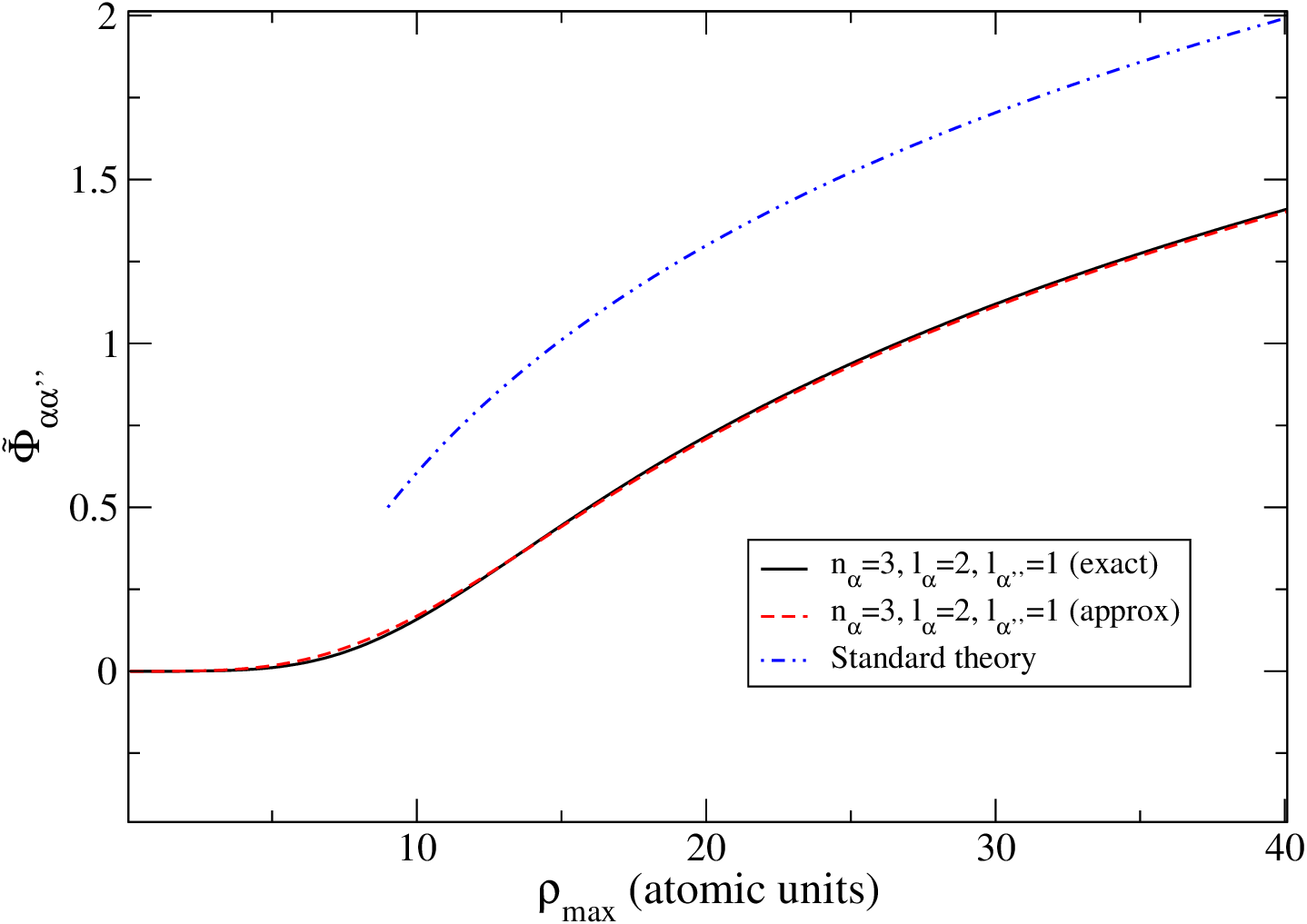}
\end{center}
\caption{Comparison between the exact computation of the quantity $\tilde{\Phi}_{\alpha\alpha''}$ entering the diagonal part of the collision operator (see Eq. (\ref{diag})) and the approximate expression (\ref{diagapp}) for $n_{\alpha}=3$, $\ell_{\alpha}=2$ and $\ell_{\alpha''}=1$.}\label{fig6}
\end{figure}

\begin{figure}
\begin{center}
\vspace{1cm}
\includegraphics[width=8cm]{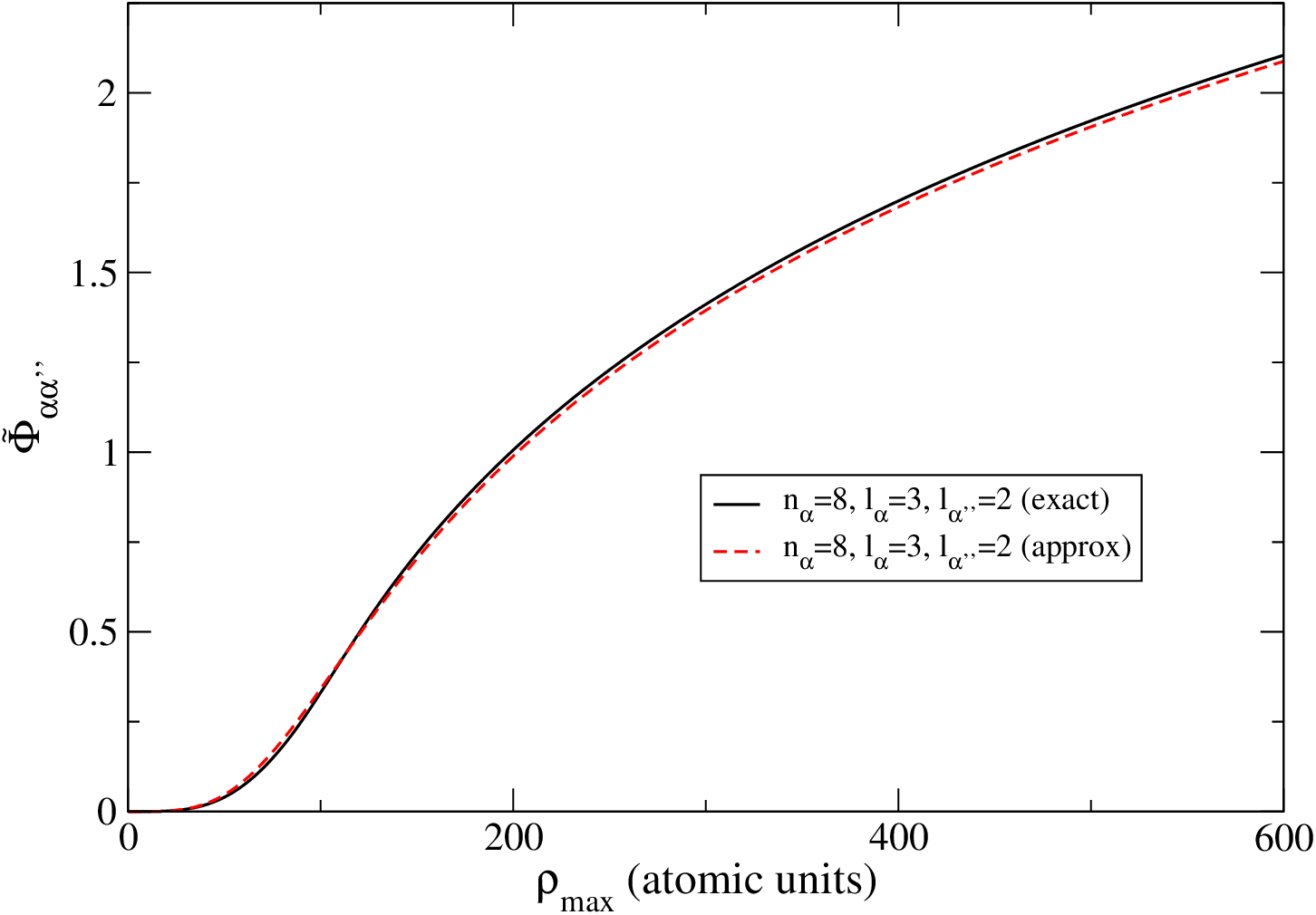}
\end{center}
\caption{Comparison between the exact computation of the quantity $\tilde{\Phi}_{\alpha\alpha''}$ entering the diagonal part of the collision operator (see Eq. (\ref{diag})) and the approximate expression (\ref{diagapp}) for $n_{\alpha}=8$, $\ell_{\alpha}=3$ and $\ell_{\alpha''}=2$.}\label{fig7}
\end{figure}

As can be checked for the two examples shown in figures \ref{fig6} and \ref{fig7}, the agreement between formula (\ref{diagapp}) and the exact results is very satisfactory. In the first case for instance, the relative error is of the order of 10 \% for $\rho_{\mathrm{max}}$=2 (but the values are very small) and less that 0.5 \% for $\rho_{\mathrm{max}}$=10. 

Expression (\ref{diagapp}) is easy to compute and facilitates the study and the accounting for penetrating collisions. It is interesting to see that the function $f$ behaves like $\ln\left(\rho_{\mathrm{max}}\right)$ (as in the standard theory without penetration effects) for high-enough values of the upper cutoff $\rho_{\mathrm{max}}$. Since the penetration standard theory is convergent for impact parameters as low as zero, there is no need for a minimum cutoff $\rho_{\mathrm{min}}$ (even though cutoffs on $v$ and $\rho$ should be introduced normally to avoid a violation of the perturbation theory, see below). In Fig. \ref{fig6}, we show also a comparison with the standard theory formula \cite{GRIEM74}:

\begin{equation}
\frac{1}{2}+\ln\left[\frac{\rho_{\mathrm{max}}}{\rho_{\mathrm{min}}}\right],
\end{equation}

\noindent where the lower cutoff $\rho_{\mathrm{min}}=n_{\alpha}^2a_0/Z$ has been introduced to avoid divergence of the logarithm. The coefficient $1/2$ represents the contribution of strong collisions in the range $0\leq\rho\leq\rho_{\mathrm{min}}$, as evaluated using the Lorentz-Weisskopf approach \cite{GRIEM59}.

It is important to mention that complication due to the cutoff in the velocity integration has been ignored in the present work. This explains the factor

\begin{equation}
\int_0^{\infty}\frac{f(v)}{v}dv=\sqrt{\frac{2m}{\pi k_BT}}
\end{equation}

\noindent in Eq. (\ref{diagapp}). As pointed out by Alexiou and Poqu\'erusse \cite{ALEXIOU05}, a velocity-dependent impact parameter $\rho_{\mathrm{min}}(v)$ is not simple to handle in the penetration theory, because to a given value of $v$ correspond(s) 0, 1 or 2 values of $\rho_{\mathrm{min}}$. On the other hand, it is more convenient to define a minimum velocity $v_{\mathrm{min}}(\rho)$ depending on the impact parameter. The role of such a cutoff for the lower limit in the velocity integration is to avoid non-unitarity issues (violation of the perturbation theory) and subsequently a strong collision term should be added in order to replace their phase-space contribution. 

The determination of $v_{\mathrm{min}}(\rho)$ is rather tedious; however, the use of the Gaussian approximation may ease this task. The idea consists in replacing the $\Delta(b)$ functions in Eq. (28) of Ref. \cite{ALEXIOU05} (which depend on the channels $\ell\rightarrow\ell\pm 1$) by an effective Gaussian function

\begin{equation}
\tilde{\Delta}\left(b\right)=\exp\left[-\frac{b^2}{2\tilde{\chi}_n^2}\right]
\end{equation}

\noindent characterized by an average parameter

\begin{equation}
\tilde{\chi}_{n}=\frac{1}{n-1}\sum_{\ell=0}^{n-2}\chi_{n,\ell,\ell+1}=\sqrt{\frac{\pi}{2}}\frac{(28n^2+n+6)}{24n}
\end{equation}

\noindent allowing us to take, if $n_a$ and $n_b$ are the principal quantum numbers of the lower and upper levels of the transition respectively:

\begin{equation}
v_{\mathrm{min}}(\rho)\approx\frac{\hbar}{m\rho}\left\{n_a^2\left(1-\exp\left[-\frac{2\rho^2}{n_a^2a_0^2\tilde{\chi}^2_{n_a}}\right]\right)-n_b^2\left(1-\exp\left[-\frac{2\rho^2}{n_b^2a_0^2\tilde{\chi}^2_{n_b}}\right]\right)\right\}.
\end{equation}

\noindent If $\rho\gg\rho_c$, one recovers the value of the standard theory:

\begin{equation}
v_{\mathrm{min}}(\rho)\rightarrow\frac{\hbar}{m\rho}\left(n_a^2-n_b^2\right).
\end{equation}

\section{Conclusion}

Alexiou and Poqu\'erusse developed a very efficient model for the effect of penetrating collisions on isolated lines of hydrogen-like ions. Their formalism includes a particular type of integrals, that the authors proposed to evaluate from a recursion relation. In the present work, we have shown that such integrals can be expressed analytically, as a finite sum involving binomial coefficients and modified Bessel functions of the third kind, which order is always smaller than the one of the function $F_q$. The exact analytical expression is easier to handle than a recursion relation, and enabled us to derive an approximate expression for the collision operator, which is very simple to enforce and accurate. Such a formula should also help to improve the understanding of strong collisions and the limits of standard theory. We also provided alternative expressions for the coefficients $a_p$ involved in the parameters entering Alexiou and Poqu\'erusse's recursion relations, and analytical expressions for integrals entering the penetration coefficient $C_{\lambda}$. 

\appendix

\section{Calculation of integrals in the penetration coefficient $C_{\lambda}$}\label{clambda}

The quantity $C_{\lambda}$ is a factor accounting exactly for penetration \cite{ALEXIOU05,POQUERUSSE06,ALEXIOU17}:

\begin{equation}\label{cl0}
C_{n,\ell,n',\ell';\lambda}(R)=\frac{\int_0^{R}P_{n\ell}(r)P_{n'\ell'}(r)r^{\lambda}dr}{\int_0^{\infty}P_{n\ell}(r)P_{n'\ell'}(r)r^{\lambda}dr}+R^{2\lambda+1}\frac{\int_{R}^{\infty}P_{n\ell}(r)P_{n'\ell'}(r)r^{-(\lambda+1)}dr}{\int_0^{\infty}P_{n\ell}(r)P_{n'\ell'}(r)r^{\lambda}dr},
\end{equation}

\noindent where $P_{n\ell}(r)$ is the radial part of the wavefunction, $R$ is the position of the perturber at time $t$ and $\lambda$ is the multipolarity ($\lambda=1$ corresponds to dipole and $\lambda=2$ to quadrupole). $n$ and $\ell$ are respectively the principal and orbital quantum numbers of the upper level and $n'$ and $\ell'$ the principal and orbital quantum numbers of the lower level. In any case, $C_{\lambda}$ can be put in the form \cite{ALEXIOU05,POQUERUSSE06,ALEXIOU17}:

\begin{equation}
C_{\lambda}(u)=1-e^{-aR}\mathcal{P}_{2n+\lambda}\left(aR\right),
\end{equation}

\noindent with $\mathcal{P}_i$ a polynomial of order $i$.

In the case of the hydrogen atom (in the following, we set $n=n'$), the expression of $P_{n\ell}(r)$ is 

\begin{equation}
P_{n\ell}(r)=-\left[\frac{Z}{n^2a_0}\frac{(n-\ell-1)!}{(n+\ell)!^3}\right]^{1/2}e^{-\frac{Zr}{na_0}}\left(\frac{2Zr}{na_0}\right)^{\ell+1}\mathcal{L}_{n+\ell}^{2\ell+1}\left(\frac{2Zr}{na_0}\right).
\end{equation}

Equation (\ref{cl0}) becomes (for clarity we use in the following the notation $C_{\lambda}(R)$ \emph{i.e.} we do not mention the quantum numbers in the subscript anymore):

\begin{equation}\label{cl}
C_{\lambda}(R)=\frac{\int_0^{aR}e^{-z}z^{\lambda+\ell+\ell'+2}\mathcal{L}_{n+\ell}^{2\ell+1}(z)\mathcal{L}_{n+\ell'}^{2\ell'+1}(z)dz}{\int_0^{\infty}e^{-z}z^{\lambda+\ell+\ell'+2}\mathcal{L}_{n+\ell}^{2\ell+1}(z)\mathcal{L}_{n+\ell'}^{2\ell'+1}(z)dz}+\left(aR\right)^{2\lambda+1}\frac{\int_{aR}^{\infty}e^{-z}z^{-\lambda+\ell+\ell'+1}\mathcal{L}_{n+\ell}^{2\ell+1}(z)\mathcal{L}_{n+\ell'}^{2\ell'+1}(z)dz}{\int_0^{\infty}e^{-z}z^{\lambda+\ell+\ell'+2}\mathcal{L}_{n+\ell}^{2\ell+1}(z)\mathcal{L}_{n+\ell'}^{2\ell'+1}(z)dz},
\end{equation}

\noindent where $a=2Z/\left(na_0\right)$ and $\mathcal{L}_p^q$ corresponds to the associated (or generalized) Laguerre polynomial in the convention of Sakurai \cite{SAKURAI93}, which is different from the one of Ref. \cite{ABRAMOWITZ64} $L_p^q$. We have

\begin{equation}
\mathcal{L}_{p+q}^q(x)=(-1)^q(p+q)!~L_p^q(x).
\end{equation}

\noindent For instance, in the present case, $p=n-\ell-1$ and $q=2\ell+1$ yield

\begin{equation}
\mathcal{L}_{n+\ell}^{2\ell+1}(x)=(-1)^{2\ell+1}(n+\ell)!~L_{n-\ell-1}^{2\ell+1}(x).
\end{equation}

The coefficient $C_{\lambda}(R)$ was shown by Alexiou and Poqu\'erusse \cite{ALEXIOU05} to be equal to:

\begin{equation}
C_{\lambda}(R)=1-\frac{e^{-aR}}{D}\sum_{k=\lambda+\ell+\ell'+2}^{2n+\lambda}c_k\left[k!\sum_{r=0}^k\frac{\left(aR\right)^{k-r}}{(k-r)!}-(k-2\lambda-1)!\sum_{r=0}^{k-2\lambda-1}\frac{\left(aR\right)^{k-r}}{(k-r-2\lambda-1)!}\right],
\end{equation}

\noindent where

\begin{equation}
D=\sum_{k=\lambda+\ell+\ell'+2}^{2n+\lambda}c_kk!.
\end{equation}

The coefficient $c_k$ is given by

\begin{equation}
c_k=a_{k-\lambda-\ell-\ell'-2},
\end{equation}

\noindent where $a_p$ is the coefficient of $z^p$ in

\begin{equation}\label{prod}
\mathcal{L}_{n+\ell}^{2\ell+1}(z)\mathcal{L}_{n+\ell'}^{2\ell'+1}(z)=\sum_{p=0}^{2n-\ell-\ell'-2}a_pz^p.
\end{equation}

In order to obtain $a_p$, Alexiou and Poqu\'erusse considered the polynomial expansion of generalized Laguerre polynomials:

\begin{equation}\label{sak}
\mathcal{L}_{n+\ell}^{2\ell+1}(z)=\sum_{k=0}^{n-\ell-1}(-1)^{k+2\ell+1}\frac{\left[(n+\ell)!\right]^2}{(n-\ell-1-k)!(2\ell+1+k)!k!}z^k.
\end{equation}

Combining Eqs. (\ref{prod}) and (\ref{sak}), we get

\begin{eqnarray}\label{ap}
a_p&=&(-1)^p\left[(n+\ell)!(n+\ell')!\right]^2\sum_{\max(0,p+1-n+\ell')}^{\min(n-\ell-1,p)}\left[(n-\ell-1-k)!\right.\nonumber\\
& &\times\left.(2\ell+1+k)!~k!~(n-\ell'-1-p+k)!(2\ell'+1+p-k)!(p-k)!\right]^{-1}.
\end{eqnarray}

However, we found that it is possible to obtain an expression, which is not a reformulation of Eq. (\ref{ap}), in terms of binomial coefficients. We can write

\begin{equation}
\mathcal{L}_{n+\ell}^{2\ell+1}(z)=(n+\ell)!\sum_{k=0}^{n-\ell-1}\frac{(-1)^{k+1}}{k!}\bin{n+\ell}{2\ell+1+k}z^k,
\end{equation}
 
\noindent yielding

\begin{equation}
a_p=\frac{(-1)^{p}}{p!}(n+\ell)!(n+\ell')!\sum_{r=\max(0,p-n-\ell')}^{\min(p,n+\ell)}\bin{p}{r}\bin{n+3\ell'+1}{n+\ell'-p+r}\bin{n+3\ell+1}{n+\ell-r}.
\end{equation}

In the following, we show that the integrals entering $C_{\lambda}(R)$ can be obtained from the generating function of Laguerre polynomials \cite{BRANSDEN83}. Let us first put Eq. (\ref{cl}) in the form

\begin{equation}
C_{\lambda}(R)=\frac{G_{n+\ell}^{n+\ell'}(\lambda,aR)}{G_{n+\ell}^{n+\ell'}(\lambda,\infty)}+\left(aR\right)^{2\lambda+1}\frac{G_{n+\ell}^{n+\ell'}(-\lambda-1,\infty)-G_{n+\ell}^{n+\ell'}(-\lambda-1,aR)}{G_{n+\ell}^{n+\ell'}(\lambda,\infty)},
\end{equation}

\noindent where 

\begin{equation}
G_{k_1}^{k_2}(\lambda,w)=\int_0^{w}e^{-z}z^{\lambda+\ell+\ell'+2}\mathcal{L}_{k_1}^{2\ell+1}(z)\mathcal{L}_{k_2}^{2\ell'+1}(z)dz.
\end{equation}

The generating function for associated Laguerre polynomials is

\begin{equation}
U_{2\ell+1}(z,s)=\frac{(-s)^{2\ell+1}e^ {-zs/(1-s)}}{(1-s)^{2\ell+2}}=\sum_{i=0}^{\infty}\frac{\mathcal{L}_{2\ell+1+i}^{2\ell+1}(z)}{(2\ell+1+i)!}s^{2\ell+1+i}
\end{equation}

\noindent and one obtains

\begin{equation}\label{ex}
\int_0^{w}e^{-z}z^{\lambda+\ell+\ell'+2}U_{2\ell+1}(z,s)U_{2\ell'+1}(z,t)dz=\sum_{i=0}^{\infty}\sum_{j=0}^{\infty}\frac{s^{2\ell+1+i}t^{2\ell'+1+j}}{(2\ell+1+i)!(2\ell'+1+j)!}G_{2\ell+1+i}^{2\ell'+1+j}(\lambda,w).
\end{equation}

The integral on the left-hand side of Eq. (\ref{ex}) is equal to

\begin{equation}\label{mgs}
\int_0^{w}e^{-z}z^{\lambda+\ell+\ell'+2}U_{2\ell+1}(z,s)U_{2\ell'+1}(z,t)dz=\frac{s^{2\ell+1}t^{2\ell'+1}}{(1-s)^{2\ell+2}(1-t)^{2\ell'+2}}\frac{\Gamma(\lambda+\ell+\ell'+3,wf(s,t))}{\left[f(s,t)\right]^{\lambda+\ell+\ell'+3}},
\end{equation}

\noindent where 

\begin{equation}
f(s,t)=1+\frac{s}{1-s}+\frac{t}{1-t}
\end{equation}

\noindent and $\Gamma(b,x)$ is the incomplete Gamma function

\begin{equation}
\Gamma(b,x)=\int_0^xe^{-z}z^{b-1}dz.
\end{equation}

Eq. (\ref{mgs}) reads also

\begin{eqnarray}\label{mgs2}
\int_0^{w}e^{-z}z^{\lambda+\ell+\ell'+2}U_{2\ell+1}(z,s)U_{2\ell'+1}(z,t)dz&=&\Gamma\left[\lambda+\ell+\ell'+3,w\frac{(1-st)}{(1-s)(1-t)}\right]\nonumber\\
& &\frac{s^{2\ell+1}t^{2\ell'+1}(1-s)^{\lambda+\ell'-\ell+1}(1-t)^{\lambda+\ell-\ell'+1}}{\left(1-st\right)^{\lambda+\ell+\ell'+3}}.
\end{eqnarray}

Therefore, in order to obtain the integrals of Eq. (\ref{cl}), we have to expand the right-hand side of Eq. (\ref{mgs}) (or Eq. (\ref{mgs2})) as a power series of $s$ and $t$, and to identify the coefficient of $s^{n+\ell}t^{n+\ell'}$ with the one provided by the right-hand side of Eq. (\ref{ex}) (\emph{i.e.} setting $i=n+\ell-2\ell-1=n-\ell-1$ and $j=n+\ell'-2\ell'-1=n-\ell'-1$) for $w=aR$ and $w\rightarrow\infty$. In the latter case, the incomplete Gamma function reduces to a usual Gamma function and we get

\begin{eqnarray}
G_{n+\ell}^{n+\ell'}(\lambda,\infty)&=&(-1)^{\ell+\ell'}(n+\ell)!(n+\ell')!(\lambda+\ell+\ell'+2)!\nonumber\\
& &\times\sum_{i=i_{\mathrm{min}}}^{i_{\mathrm{max}}}\bin{\lambda+n+\ell'+1-i}{n-\ell-1-i}\bin{\lambda+\ell'-\ell+1}{i}\bin{\lambda+\ell-\ell'+1}{\ell-\ell'+i},
\end{eqnarray}

\noindent with $i_{\mathrm{min}}=\max(0,\ell'-\ell)$ and $i_{\mathrm{max}}=\min(n-\ell-1,\ell'-\ell+\lambda+1,\lambda+1)$, which is a particular case of

\begin{eqnarray}\label{new}
\int_0^{\infty}e^{-z}z^{\alpha}\mathcal{L}_{m_1}^{k_1}(z)\mathcal{L}_{m_2}^{k_2}(z)dz&=&(-1)^{m_1+m_2}m_1!m_2!\alpha!\sum_{i=i_{\mathrm{min}}}^{i_{\mathrm{max}}}\bin{\alpha+m_1-k_1-i}{m_1-k_1-i}\bin{\alpha-k_1}{i}\nonumber\\
& &\times\bin{\alpha-k_2}{m_2-k_2-m_1+k_1+i},
\end{eqnarray}

\noindent with $i_{\mathrm{min}}=\max(0,m_1-k_1-m_2+k_2)$ and $i_{\mathrm{max}}=\min(m_1-k_1,\alpha-k_1,\alpha-m_2+m_1-k_1)$. It is worth mentioning that recurrence relations exist (see for instance Ref. \cite{BLANCHARD74}) as well as explicit expressions (see for example Refs. \cite{SHERTZER91,MARXER95}). Expression (\ref{new}) can be useful for many applications, through the calculation of expectation values $\langle n\ell|r^j|n\ell'\rangle$, where $t$ is an integer:

\begin{equation}
\int_0^{\infty}r^jP_{n\ell}P_{n\ell'}(r)dr.
\end{equation}

For the incomplete integral (finite value of $w$), we have to expand also $\Gamma\left[\lambda+\ell+\ell'+3,w\frac{(1-st)}{(1-s)(1-t)}\right]$ in powers of $s$ and $t$. This can be done using \cite{ABRAMOWITZ64}:

\begin{equation}
\Gamma(b,x)=\sum_{r=0}^{\infty}\frac{(-1)^rx^{b+r}}{(b+r)r!}
\end{equation}

\noindent and

\begin{equation}
\left(1+\frac{s}{1-s}+\frac{t}{1-t}\right)^r=\sum_{k=0}^{r}\sum_{i=k}^{\infty}\sum_{j=k}^{\infty}\alpha_{ijk}(r)s^{i}t^{j},
\end{equation}

\noindent with

\begin{equation}
\alpha_{ijk}(r)=(-1)^k\bin{r}{k}\bin{r+i-k-1}{r-1}\bin{r+j-k-1}{r-1}.
\end{equation}

The calculation is more complicated and the result less compact than in the case $w\rightarrow\infty$. The integrals involved in the coefficient $C_{\lambda}$ can also be simplified using the Feldheim formula \cite{FELDHEIM40,POPOV03}, which expresses the product of two generalized Laguerre polynomials as a linear combination of generalized Laguerre polynomials.



\end{document}